\documentclass{mn2e}

\usepackage[dvips]{graphicx}

\def\etal{et al.}

\begin{document}

\def\etal{et al.}


\title[]{Global environmental effects versus galaxy interactions}
\author[Perez et al.]{Josefa Perez$^{1,2,3}$\thanks{E-mail:
jperez@fcaglp.unlp.edu.ar}, Patricia Tissera$^{1,3}$, Nelson Padilla$^{4}$, M. Sol Alonso$^{3,5}$ and Diego G. Lambas $^{3,6}$\\
$^{1}$Instituto de Astronom\'\i a y F\'\i sica del Espacio,Conicet-UBA, CC67, Suc.28,Ciudad de  Buenos Aires, Argentina.\\
$^{2}$Facultad de Ciencias Astronom\'\i a y Geof\'\i sica, Universidad Nacional de La Plata, Argentina.\\
$^{3}$Consejo Nacional de Investigaciones Cient\'\i ficas y T\'ecnicas, CONICET, Argentina. \\
$^{4}$Departamento de Astronom\'\i a y Astrof\'\i sica, Pontificia Universidad Cat\'olica de Chile, Santiago, Chile.\\
$^{5}$Complejo Astron\'omico El Leoncito, CP J5402DSP, San Juan, Argentina.\\
$^{6}$Observatorio Astron\'omico de la Universidad Nacional de C\'ordoba, Argentina.}


\maketitle

\begin{abstract}

We explore  properties of close galaxy pairs and merging systems  selected
 from the SDSS-DR4 in different  environments with the aim to assess the relative
importance of the role of 
 interactions over global environmental processes. 
For this purpose, we 
perform a  comparative study of galaxies with and without close companions
 as a function of local density and host-halo mass, carefully removing
sources of possible biases.
 
We find that at low and high local density environments, colours and morphologies 
of close galaxy pairs are very similar to those of isolated galaxies. At intermediate densities, 
we detect significant differences, indicating that close pairs could have experienced a more rapid transition onto the red sequence
than isolated galaxies.
 The presence of a correlation between colours and morphologies indicates that the physical mechanism responsible 
for the colour transformation also operates changing galaxy morphologies.
At fixed local densities, we find a dependence of the red galaxy fraction on dark matter halo mass   for galaxies with or without a close companion. This suggests the action of host halo mass related effects.  
Regardless of dark matter halo mass, 
we  show that  the percentage of red galaxies in close pairs and in the control sample are comparable
at  low and high local density environments.
However, at  intermediate local densities, the gap in the red fraction between close pairs and the control galaxies  increases from  $\sim 10\%$ in low mass haloes up to $\sim 50\%$ in the most massive ones. 
 Interestingly, we also detect
that $ 50\%$ of merging systems populate the intermediate local environments, with a large fraction of them being extremely red and bulge dominated. 
Our findings suggest that in intermediate density environments 
 galaxies are efficiently pre-processed  by close encounters and mergers before entering higher
local density regions.
\end{abstract}


\section{Introduction}

Several observational and theoretical works have gathered evidence to determine that the
 environment where galaxies reside plays a fundamental role in shaping
their properties. However, although  the transformation of 
blue, late-type and star-forming field galaxies
 into red, early-type and passive cluster galaxies have been well established 
(Oemler 1974; Dressler 1980; Lewis et al. 2002; Gomez et al. 2003; Balogh et al. 2004; Baldry et al. 2004;
O'Mill, Padilla \& Lambas, 2008), 
there is no consensus on the mechanisms responsible for this transformation. Many explanations
have been proposed  including:  i) {\em  ram-pressure stripping} of  cold interstellar gas of 
galaxies falling at high velocities into the ICM, which produces a fast truncation of the star formation
(Gunn \& Gott 1972); ii) {\em starvation} or 
{\em strangulation}, which are also stripping gas processes of the hot diffuse component 
of satellite galaxies,  which  affects the star formation on longer timescales (Larson et al. 1980);   
iii) {\em harassment}, the cumulative effect of several rapid encounters with other 
cluster members, which leads to substantial changes in the galaxy morphology  (Moore et al. 1998);   
iv) {\em mergers} and {\em interactions} of galaxies which can trigger an intense burst of star formation, 
rapidly consuming the cold gas and forming spheroidal systems 
(Toomre \& Toomre 1972, Kauffmann et al. 1993). 
These explanations, however, are still under discussion.

van den Bosch et al. (2008) and Weinmann et al. (2008) analyse the role of
satellite quenching for the build-up the red galaxy sequence. They find that the environmental
processes  which shut down the star formation (SF) activity in satellite galaxies are equally efficient in host haloes
of all masses.  This  rules against mechanisms that are thought to operate
only in very massive haloes, such as ram-pressure or harassment.   
They suggest that the process responsible for quenching the SF in satellites should last a timescale of a few Gyr, 
suggesting starvation as the satellite-specific transformation mechanism. 
They also claim that an additional mechanism is also required  
because quenching alone cannot explain the morphological transformations in 
the build-up of the red sequence.

Alternatively, attempts have been made to remove the problem 
from clusters entirely by proposing a preprocessing of disc blue galaxies 
to red earlier type systems at moderate environments
(Balogh et al. 2004; Mihos 2004; Moss 2006; Patel et al. 2008). Recently, many authors
have investigated the dependence of galaxy properties on environment at intermediate
densities (i.e. galaxy groups in the outskirt of clusters,  infall populations) suggesting
different mechanisms to account for them. 
Patel et al. (2008) find that galaxies at cluster-centric radii
larger than 3 Mpc show an enhanced red galaxy fraction, indicating that
intermediate density regions and groups in the outskirts of clusters are
locations where the local environment influences the transition of galaxies onto the
red-sequence, as opposed to mechanisms that operate on cluster scales 
(e.g. ram-pressure stripping, harassment). In the same direction, Moss (2006)  provides evidence 
stating that the cluster giant S0 population can be explained 
as the outcome of minor mergers with the infalling population integrated over the past $\sim 10$ Gyr. 

Analysis of the  SF at intermediate environments show that the current SFR of a galaxy  
falling along a supercluster filament is likely to undergo 
a sudden enhancement before the galaxy reaches the virial radius of the
cluster (Porter et al. 2008). These authors suggest that the main process responsible
for this rapid burst are close interactions with other galaxies 
in the same filament, if the interactions occur before the gas reservoir
of the galaxy gets stripped off due to the interaction with the ICM. 

On the other hand, Gallazzi et al. (2008) explore the amount of obscured SF as a function of environments 
in the A901/902 supercluster. The SF hidden among red galaxies is detected by using SF indicators 
that are not affected by dust attenuation. Otherwise, they could be missed 
or mistakenly classified as post-starburst on the basis of their weak emission lines obtained via optical, 
dust-sensitive SFR indicators. Combining the near UV/optical data  with infrared photometry, 
they find that $\sim 60\%$ of red star-forming galaxies have IR-to-UV luminosity ratios 
which indicate high dust obscuration. Interestingly, most of them populate intermediate density regions. 
In agreement with this result, Wolf et al. (2005) have also identified an excess of dusty red galaxies 
with young stellar populations  in the infalling region of the same cluster. 
More evidence of dust-obscured SF at intermediate regions is provided by Miller \& Owen (2002). They 
 find that up to  $\sim 20\%$ of  the galaxies in 20 nearby Abell clusters 
have dust-obscured SF and are preferentially located at intermediate density regions,  
 with respect to normal star-forming galaxies or AGNs.
In addition, Poggianti et al. (2008) report that dusty starburst candidates present 
a very different environmental dependence than post-starburst galaxies.
They find that the spectra of dusty starburst candidates are 
numerous in all environments at intermediate redshifts, particularly among
galaxy groups. This favours the hypothesis of dusty starbursts triggered by mergers, 
expected to be common in groups. 

Motivated by these findings, in this paper, we revise the role of 
mergers and galaxy interactions in driving galaxy evolution at different
density environments by using the SDSS-DR4 data.  In a hierarchical clustering universe, as Mihos (2004) noticed, galaxy clusters would form not by accreting individual
galaxies from the field, but rather through the infall of less massive groups moving
along the filaments. Such infalling groups provide locations with much lower velocity dispersions
than the cluster medium, thus permitting strong slow encounters more normally associated with the field. 
In agreement, Moss  (2006) shows that $\sim 50\% - 70\%$  of the infall population 
are found to be in merging systems and slow galaxy-galaxy encounters.
Hence, we will focus 
our analysis on close encounters and merger candidates in order to disentangle their role 
at such intermediate density regions. But before going any further, we analyse
a technical issue.

The effect of galaxy interactions has been largely studied in
optical and infrared surveys (Lambas et al. 2003; Nikolic et al. 2004;  Alonso et al. 2004;
De Propris et al. 2005; Kewley et al. 2006; Alonso et al. 2006; Lin et al. 2007; Michel-Dansac et al. 2008; Park et al. 2009)
 and also using numerical simulations (Barnes \& Hernquist 1996;
Mihos \& Hernquist 1996; Larson \& Tinsley 1978;  Perez et al. 2006a,b). 
These results conclude that mergers and close interactions 
could actually modify the star formation, morphology, colours and metallicities of galaxy pairs. 
In all these cases, people have attempted to isolate the
effects of galaxy interactions by comparing galaxies in  pairs with isolated galaxies. 
However, different authors have proposed different ways to build these control
samples (CS). 
 By  using SDSS-DR4 mock galaxy catalogues  built using the Millennium Simulation,
  Perez et al. (2009, hereafter P09) show that the set of constrains used to define a CS might introduce biases
which could affect the interpretation of results.  The fact that the physics of interactions is not
included in the semi-analytic model (SAM) studied in P09 allows the authors   to attribute 
 differences between the mock control and pair samples solely to  selection biases.
P09 find that these  biases are diminished by $\approx 70\%$ after
imposing constrains on redshifts, stellar masses and local densities, 
and are completely removed when halo masses are also  considered (also see Barton et al. 2007).
However, P09 notice that the contribution of the halo mass  bias introduced by
the recipes adopted in SAMs to model satellite galaxies is probably exacerbated.
Based on these previous theoretical findings, we first build  an unbiased CS from the SDSS-DR4 data 
in order to isolate signals of galaxy interactions in a more robust way.

This paper is organized  as follows. In Section 2, we describe 
the SDSS-DR4  Galaxy Pair Catalogue and CS. We  analyse possible bias effects in the 
selection of CS (2.1) and correct them in order to obtain an unbiased CS, 
suitable for isolating the effect of galaxy interactions (2.2). 
In Section 2.3, we revise some of the previous observational results of 
galaxy pairs by comparing them with isolated galaxies using  the unbiased CS. 
The role of galaxy interactions and mergers in different local density and host-halo mass environments are 
discussed in Section 3. Finally, in Section 4 we summarize our results.




\section{Building a control sample for a galaxy pair catalogue.}

The analysis of this paper is based on the SDSS-DR4  photometric and spectroscopic galaxy catalogue, for which there are estimations of 
 gas-phase oxygen abundances,  stellar masses and  metallicities provided 
by   Tremonti et al. (2004) and Gallazzi et al. (2005).
Star formation rate  estimates for galaxies  are obtained as described in Brinchmann et al. (2004).
The SDSS-DR4 galaxy sample is essentially a magnitude limited spectroscopic sample with $r_{lim}<17.77$ covering
a redshift range $0<z<0.25$.  We  considered a shorter redshift range,
$0.01<z<0.1$, in order to avoid strong incompleteness 
at larger distances (Alonso et al. 2006). We  also excluded 
from our sample AGNs, which could 
affect our interpretation of results due to contributions from their emission line spectral
features. 

We characterized the local environment of galaxies defining a projected
density parameter, $\Sigma$. This parameter is calculated by using the projected distance to the
$5^{th}$ nearest neighbour, $\Sigma= 5/(\pi d_{5}^{2})$. Following Balogh et al. (2004), 
neighbours have been chosen to have luminosities  brighter than $M_{r}<-20.5$ 
and radial velocity differences lower than $1000 \rm \,km \,s^{-1}$. In the
case of pairs, we also estimated the local density by using the 
$6^{th}$  nearest neighbour, in order to assess possible biases in the definition of the local density for galaxy pairs.  
However, we find that our results are insensitive to the definition of $\Sigma$ by using 
 either the $5^{th}$ or the  $6^{th}$ nearest neighbour.

We use the SDSS-DR4 galaxy group catalogue from Zapata et al. (2009) to assign
a host halo mass ($M_{vir}$) to each individual galaxy in our sample. This group
catalogue is complete above masses of $10^{13}$h$^{-1}M_{\odot}$, out
to the limit redshift of our galaxy sample, $z=0.1$. The mass
assignment is done by searching for the closest group, in terms of
its virial radius $r_{vir}$, to each galaxy; if a group is found
within 1.5 $r_{vir}$ in projection and with a velocity difference,
$\Delta v<1000$km/s the galaxy is assigned the group mass as its host
halo mass. Galaxies which do not satisfy these conditions for any
group are assumed to be hosted by haloes below the mass detection
limit of $10^{13}$h$^{-1}M_{\odot}$.

Following Alonso et al. (2006, and references therein), we build a Galaxy Pair Catalogue (GPC)
requiring members to have relative projected separations $r_{p}< 100 \rm \,kpc \,h^{-1}$
and relative radial velocities $\Delta V< 350 \rm \,km \,s^{-1}$. Galaxies 
without a near companion within the adopted thresholds constitute the
 Isolated Galaxy Sample (IGS). In order to properly assess the significance of
the results obtained from the GPC,  we use a control sample (CS) constructed by
selecting galaxies from the IGS. 
In a first attempt to build a CS,  we follow recent observational works  
(Lambas et al., 2003; Alonso et al., 2006; Michel-Dansac et al., 2008) and 
 define a CS by selecting   galaxies from the IGS 
which match one-to-one the redshift and r-band magnitude of each galaxy in pairs,  hereafter the SDSS-{\bf CS1}.

However, the analysis of SDSS-DR4 mock catalogues built using the Millennium Simulation shows
that  a CS defined by only applying redshift and luminosity requirements exhibits different 
stellar masses, morphologies, halo masses and local projected density distributions
than those of galaxies in pairs (P09) as discussed in the Introduction.
Hence, in this section, and based on these theoretical findings, we analyse how these biases 
could affect the selection of the SDSS-{\bf CS1}, and then we correct them
in order to build up a robust 'unbiased' control sample.

\subsection{Analysis of biases in  the selection of the SDSS-{\bf CS1}.}

In this subsection,
we apply the tests devised by P09 to the SDSS-{\bf CS1}, excluding the 
morphological selection, which we will not consider in our work since
in the case of merging systems or close galaxy pairs with tidal features, 
it is rather difficult to objectively define their morphology.

\begin{figure}
\centering
\includegraphics[width=6cm,height=5cm]{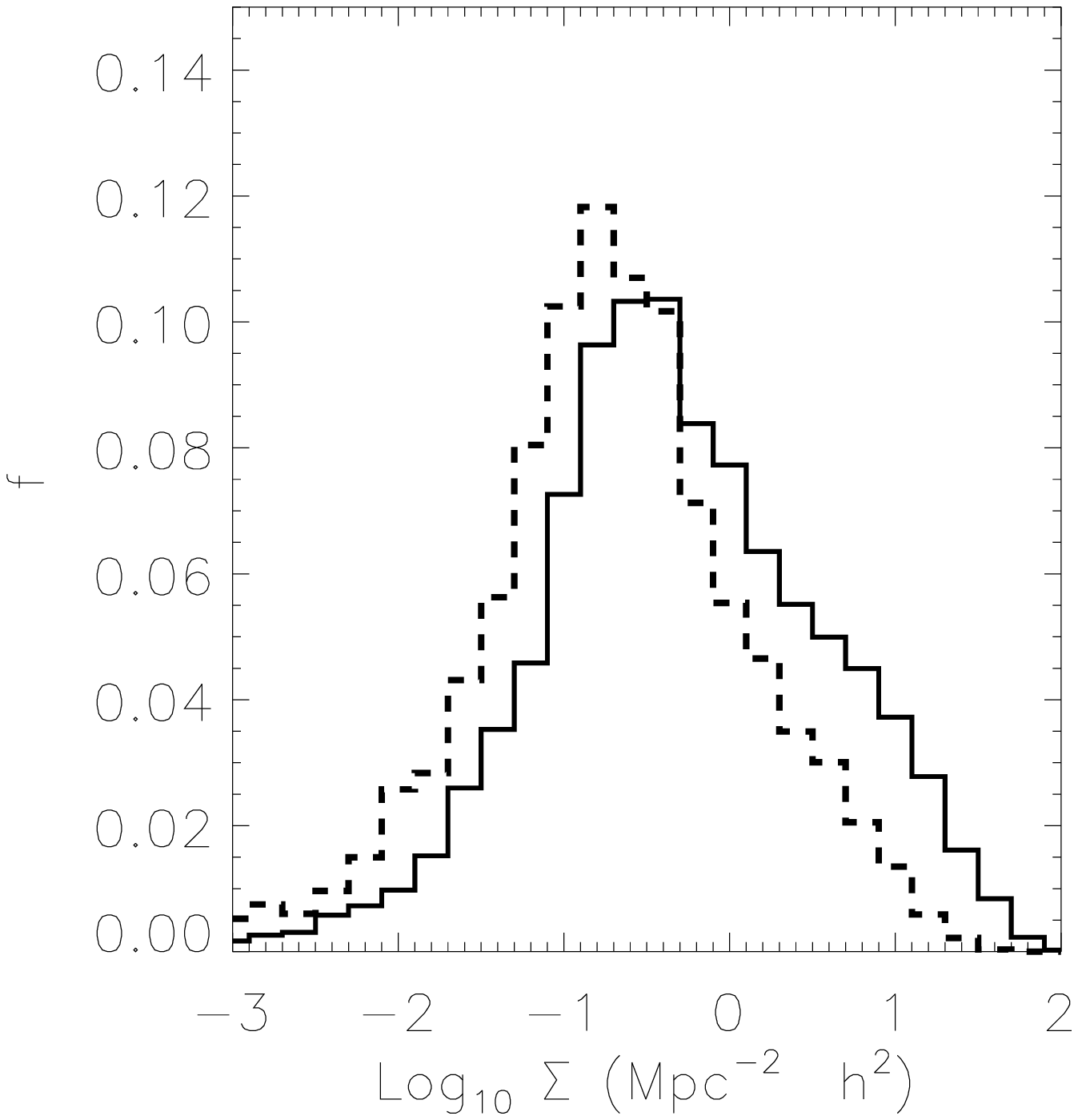}
\includegraphics[width=6cm,height=5cm]{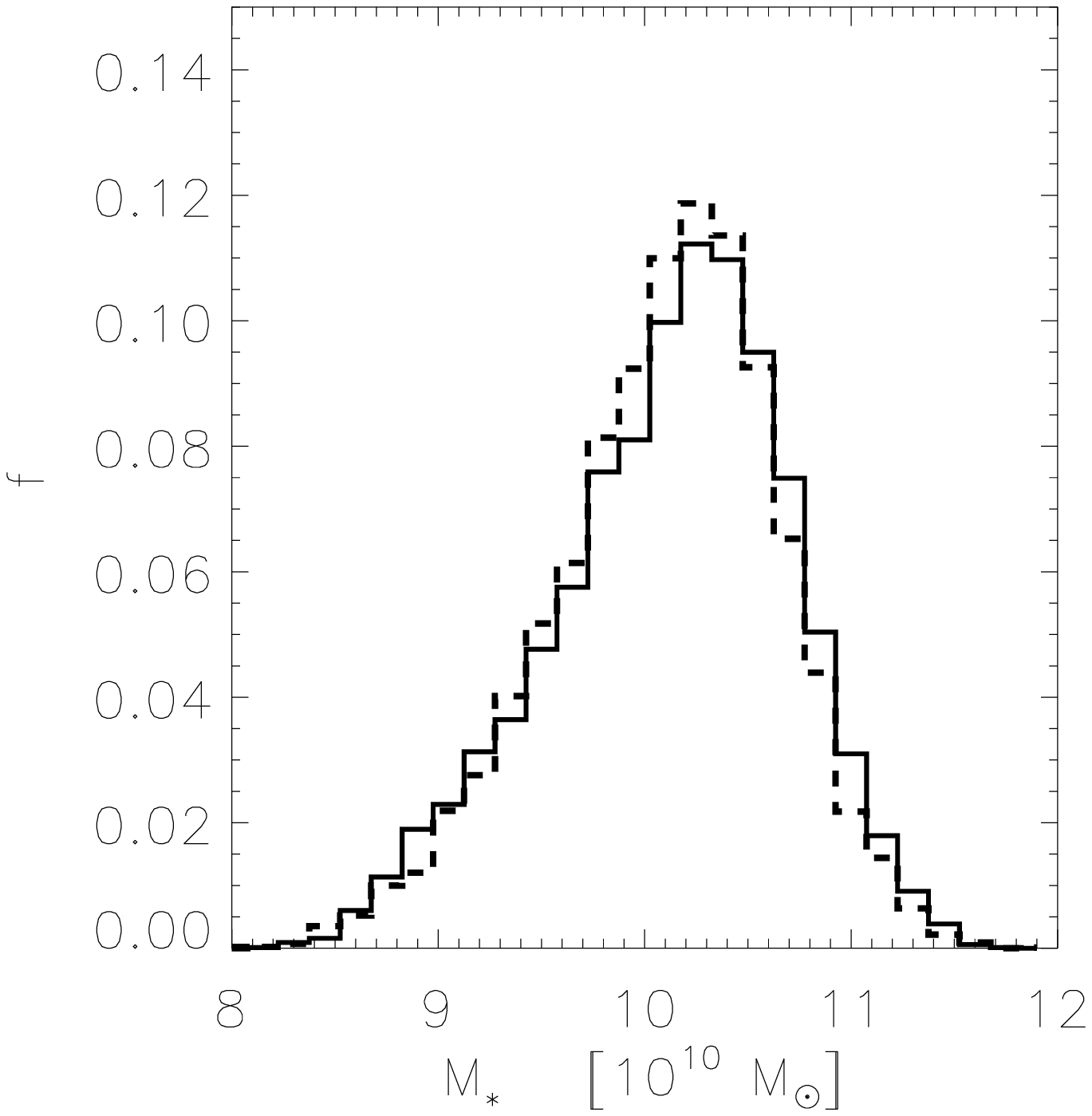}
\includegraphics[width=6cm,height=5cm]{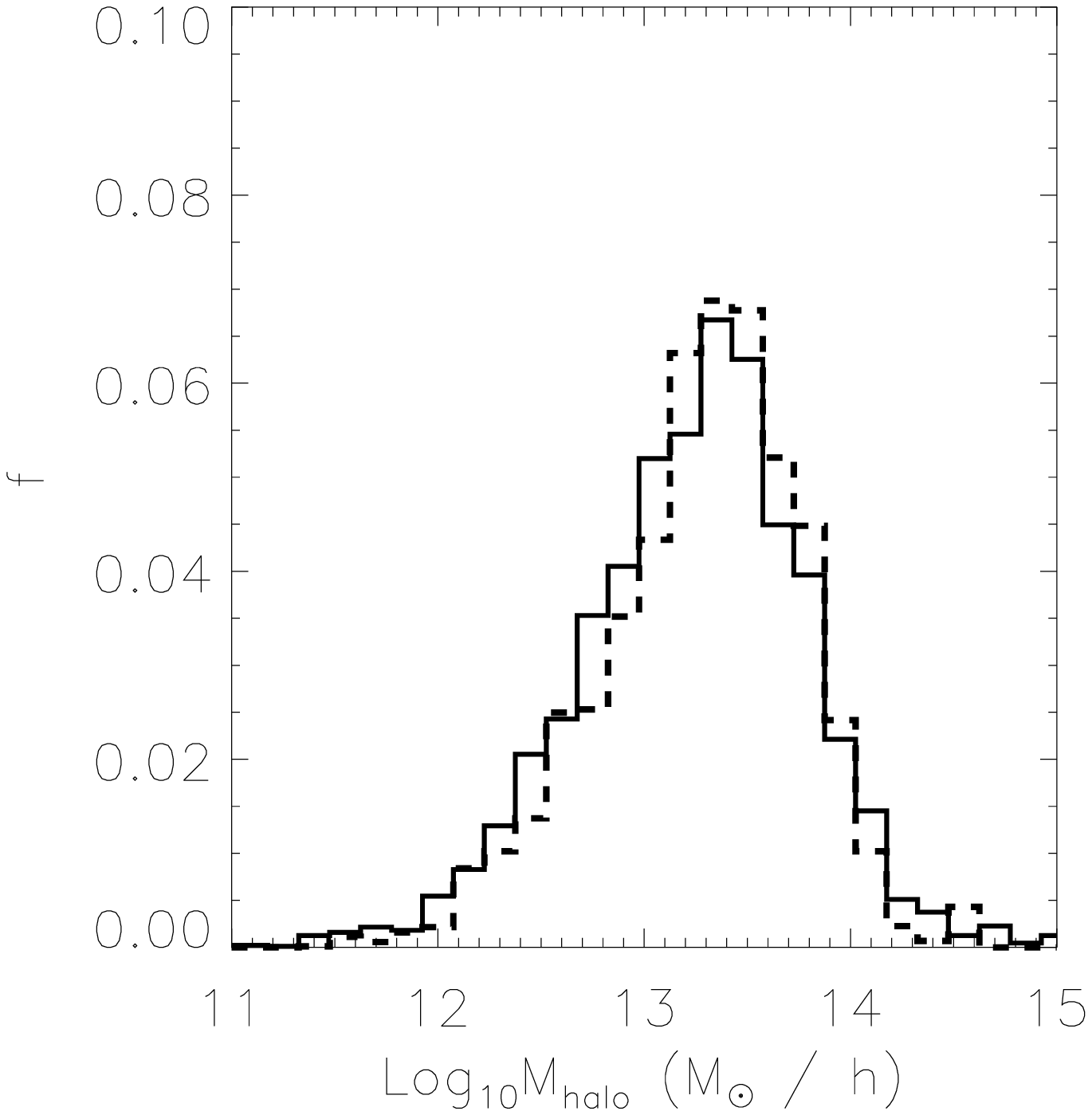}
\caption{Distributions of local density parameter ($\Sigma$) (upper panel), 
stellar masses (middle panel) and dark matter halo masses  (lower panel), 
for SDSS galaxies in pairs (solid lines) and in the SDSS-{\bf CS1} (dashed lines). }
\label{bias1}
\end{figure}

In Fig.~\ref{bias1}, we perform a comparative analysis of
 stellar masses, halo masses and local projected density ($\Sigma$) distributions  
for SDSS-DR4 galaxy pairs (solid lines) and for the SDSS-{\bf CS1} (dashed lines). 
The most significant differences between control and pair properties 
are observed for the distributions of local density environment. In agreement 
with  theoretical findings by P09, galaxy pairs tend to inhabit higher density regions than
galaxies in the SDSS-{\bf CS1}.

Consistently with  P09, we find that  galaxy pairs  
exhibit slightly larger stellar masses than their isolated counterparts in the SDSS-{\bf CS1}.
However,  observations show that 
pair and control halo mass distributions are similar, in contradiction to P09 who
reported that in  the SAM   the halo mass is 
the main factor contributing to bias the mock-{\bf CS1}.
As a matter of fact,  P09 found that
mock galaxy pairs are hosted by larger dark matter haloes than 
galaxies in the mock-{\bf CS1}. However,  
they warned that this effect could be overestimated in the SAM due to the environmental
treatment of the starvation of hot gas in satellite galaxies 
(Weinmann et al. 2006; Kang et al. 2008; P09).  Thus, the fact that SDSS galaxies with and without a near companion 
have similar halo mass distributions (Fig.~\ref{bias1}), 
could be considered as an indication of the exacerbated environmental modelling of 
satellite galaxies in the SAM. However, we remind 
the reader that the dynamical estimates of halo masses used in this present paper could  also be introducing 
potential sources of errors due to the  uncertainties introduced by velocity dispersions
  (Eke et al. 2004; Padilla et al. 2004).

The observational results shown in Fig.~\ref{bias1} indicate that even when the local density
environment is the most significant bias in the SDSS-{\bf CS1}, 
we should also take into account both stellar and halo masses  in order
to properly select the SDSS control sample.

\subsection{Treatment of selection effects}

In order to remove the bias introduced by the differences in stellar masses,  we define
 a CS using  stellar masses instead of magnitudes as a constrain. 
Thus, the  SDSS-{\bf CS2} is built up by selecting galaxies from the IGS so that the distributions of
redshifts and stellar masses match those of galaxies in pairs. 
Ellison et al. (2008) also build a CS using a similar approach to our
SDSS-{\bf CS2} in order to study
 close galaxy pairs in the SDSS.  Additionally, we build a SDSS-{\bf CS3} imposing constrains on
 redshift, stellar mass and halo mass. 
Finally,
we define the  SDSS-{\bf CS4} populated by IGS galaxies matching the galaxy pair 
redshift, stellar mass, local density environment and halo masses; the SDSS-{\bf CS4} should
not be affected by selection biases according to the results by P09.

After this CS building process, we have to remark that in order to construct the
more suitable and 'unbiased'  SDSS-{\bf CS4}, we remove 
around $8\%$  of the galaxies in pairs from the original GP sample to account for the lack of  counterparts in the IGS
with identical stellar masses, local density environments and halo masses. 
Most of the removed galaxies in pairs reside in high density regions 
where galaxies satisfying our isolation requirements are less common (Fig.~\ref{bias1}).

\subsection{Isolating the effect of interactions.}

We have shown that the set of constrains used to define the SDSS-{\bf CS1} introduces biases
which could affect the interpretation of the results found for galaxy pairs.  
In this section, we revise some previous observational analysis of galaxy pairs
from the SDSS-{\bf CS1}  (Lambas et al. 2003; Alonso et al. 2006; Michel-Dansac et al. 2008) 
 to evaluate how their results would be modified considering the different CSs defined 
in the previous subsection. 

In order to test the performance of each CS, we study their colours, metallicities and star formation activity, comparing them to those of galaxies in pairs. 
Particularly, we analyse the dependence of the star formation activity on the environment, as well as
$u-r$ colour distributions for the
SDSS-{\bf CS1}, SDSS-{\bf CS2} and SDSS-{\bf CS3},  finding similar results for all these CS (see
P09 for a discussion on the mass-metallicity relation). The most
significant change is detected when analysing the properties of SDSS-{\bf CS4}, 
indicating that, the local density environment is responsible for
introducing an important bias effect (see upper panel of Fig.1). Using results from a semi-analytical model,  P09
showed that the halo mass is the parameter with the largest contribution towards biasing a CS. 
However, they could not separate this latter bias effect from the modelling bias associated to
the treatment of satellites. The analysis of SDSS control samples shows a less
significant role of the dark matter mass in the CS definition than previously reported (P09, Barton et al. 2007), 
and contributes to show the exacerbated environmental modelling of 
satellite galaxies in the SAM (Weinmann et al. 2006; Kang et al. 2008).

\subsubsection{Comparative analysis of galaxy pairs, SDSS-{\bf CS2} and SDSS-{\bf CS4}}

Since bias effects are removed 
only after defining the SDSS-{\bf CS4}, from this point on we will only show the results for samples
SDSS-{\bf CS2} and the  'unbiased' case. 
Note that we have chosen the SDSS-{\bf CS2} instead of the SDSS-{\bf CS1}, 
since they behave similarly and, on the other hand, the stellar mass is a more fundamental quantity than the luminosity of a galaxy
(Brinchmann et al. 2000; Kauffmann et al. 2003; Panter et al. 2004). 

\begin{figure*}
\includegraphics[width=6.5cm,height=5.5cm]{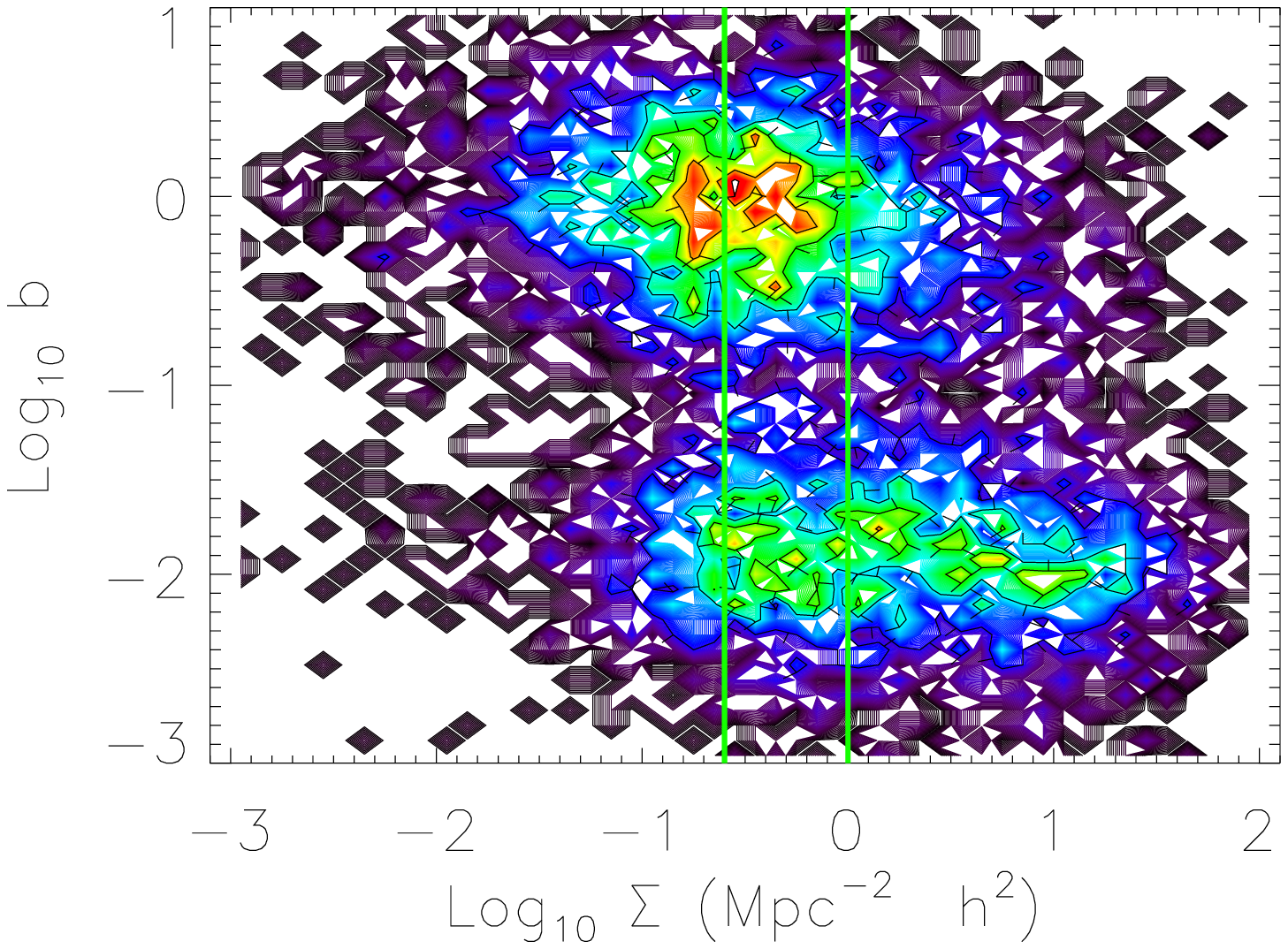}
\includegraphics[width=6.5cm,height=5.5cm]{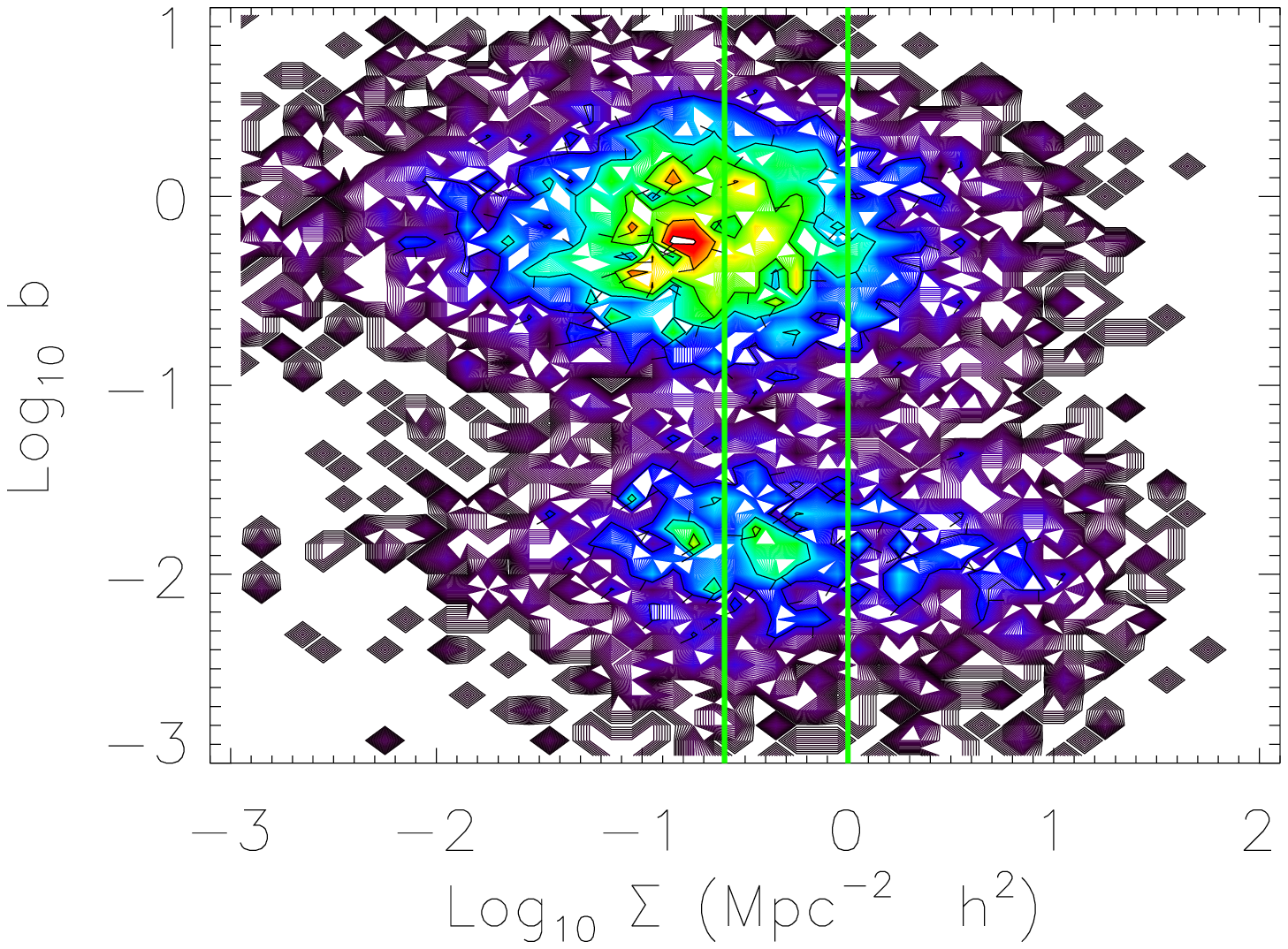}\\
\includegraphics[width=6.5cm,height=5.5cm]{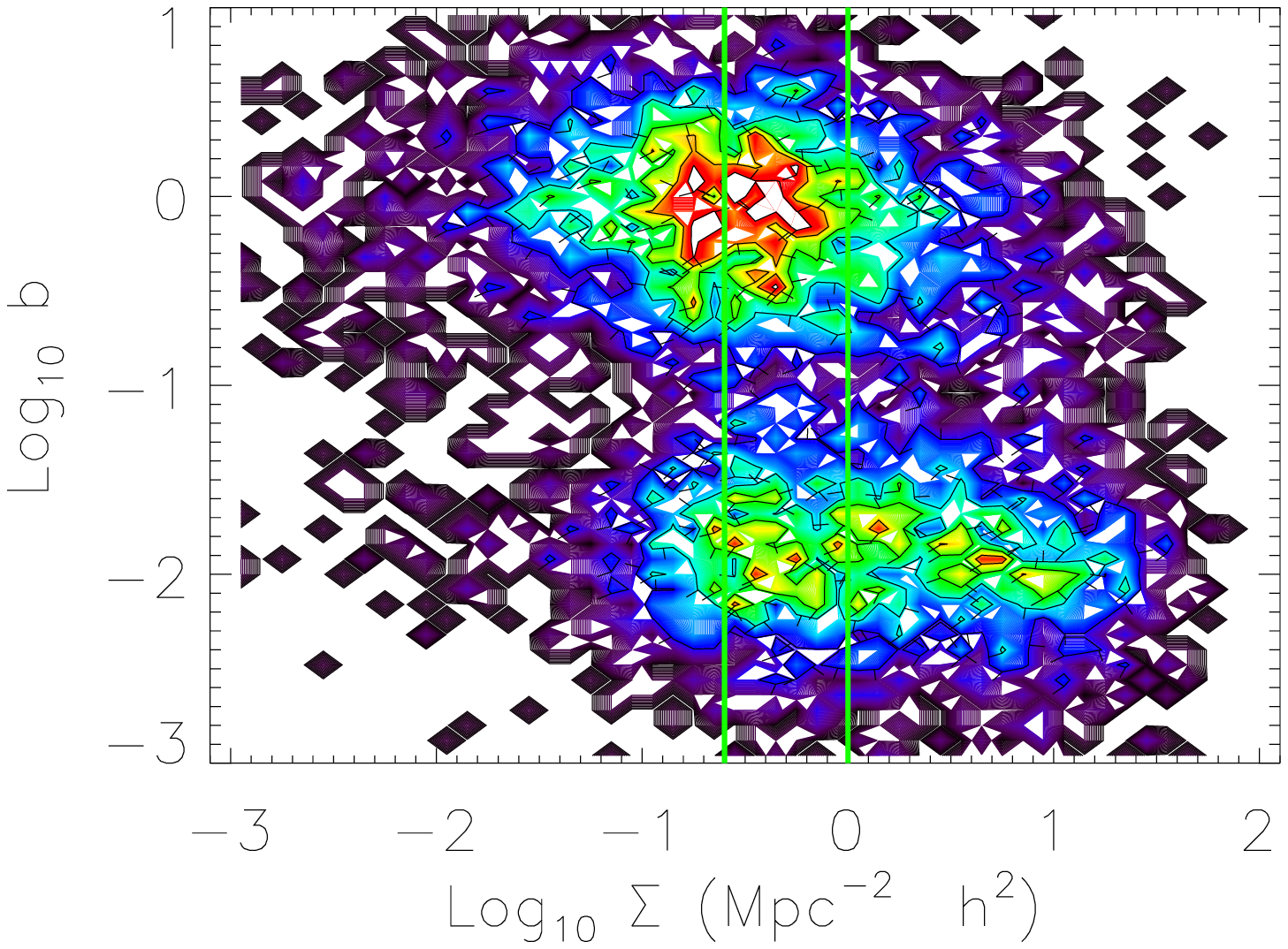}
\includegraphics[width=6.5cm,height=5.5cm]{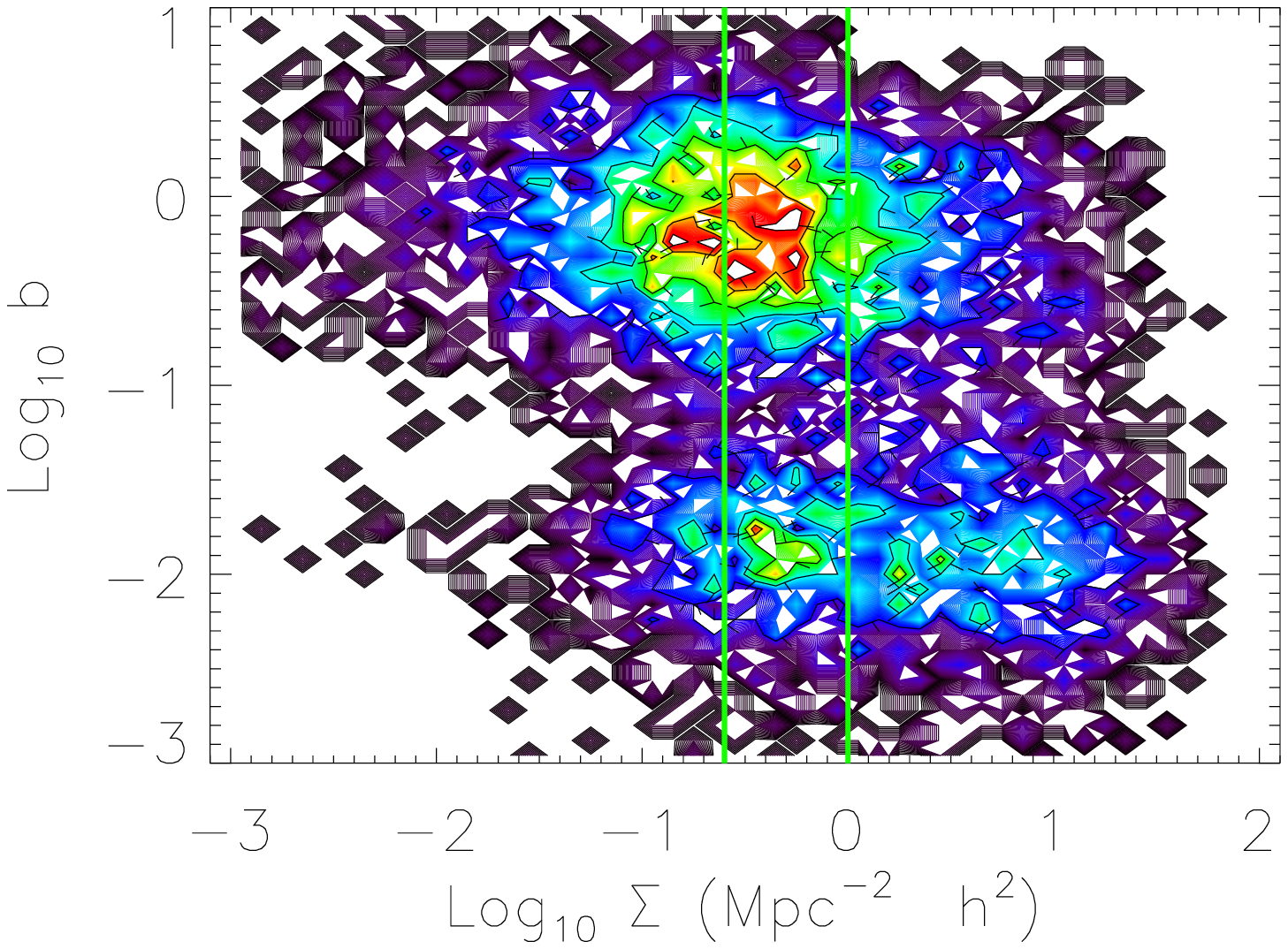}
\caption{Contour plots of the stellar birth rate  ($b$) and  local density ($\Sigma$) parameters,
for galaxy pairs (left panels) and CSs (right panels). 
Upper (lower) panels show the biased samples 2 (unbiased samples 4), see the text for more details.
The sequence from red to dark blue colours indicates a decrease in the galaxy number density. Green lines
divide low, intermediate and high density environments (see Table 1).}
\label{b-sigma}
\end{figure*}

We first analyse the SF activity in different density environments. 
Particularly, we estimate the SF history of systems by defining the stellar birthrate parameter,  
$b=0.5 \,t_{H}(SFR/M_{*})$, where $t_{H}$ is the Hubble time, and $SFR/M_{*}$ is the present star formation rate 
normalized to the total stellar mass (Brinchmann et al. 2004). 
In Fig.~\ref{b-sigma}, we show the contour plots of
the stellar birthrate parameter as a function of the local density estimator ($\Sigma$)   
for galaxy pairs (left panels)  and CSs (right panels). The upper plots show the result 
for galaxies with and without a near companion using   the 'biased' 
{\bf CS2}\footnote{For simplicity, hereafter from this point on we have suppressed  the SDSS prefix in sample names.}  definition.
Lower panels show the  'unbiased' results obtained by comparing galaxies in pairs with those in the {\bf CS4}. 
Note, that in order to define the 'unbiased' {\bf CS4}, we remove 
a small fraction of galaxies in pairs, thus the galaxy pair samples 
also change from the top to the bottom panels.  In order to clarify the discussion, from this point on we will refer to 
the galaxy pair samples as {\bf GP2} (upper left panel) and {\bf GP4} (lower left panel),
 respectively. 
The green lines in the plots divide the three different environmental regions discussed throughout this paper 
(see Table 1 for description).

As we can appreciate in the top panels of Fig.~\ref{b-sigma}, the biased samples show that while galaxies with a near 
companion (top left) have a  clearer bimodal distribution in the SF-environment relation 
(i.e. a large fraction of passive SF galaxies populating  high density regions),
isolated galaxies (top right) are more consistent with a unimodal SF distribution shifted to
high values of the stellar birthrate parameter. 
These significant differences between  galaxies in {\bf GP2} and in {\bf CS2}, however, cannot  be
only
attributed to galaxy interactions, but also to a  biased selection of these samples.
Indeed, we can see that the SF-environment relation of the 'unbiased'  {\bf CS4} (bottom right) significantly changes
with respect to that of {\bf CS2} (top right), resembling more closely that of its pair 
counterpart, {\bf GP4} (bottom left). 
This means that after correcting for the biases, 
the differences between the SF-environment distributions
of galaxies with and without a near companion are partially reduced. 
However, discrepancies still remain even when {\bf GP4} is compared to the unbiased  {\bf CS4}, 
suggesting  a real effect coming from galaxy interactions. 
We find that, at high densities, the {\bf GP4} has a larger fraction of low star-forming systems than the {\bf CS4} while 
at intermediate and low density environments, it has  larger fractions of strong star-forming galaxies.

\begin{table*}
\caption{Ranges of local density environments ($\Sigma$) and halo masses ($M_{vir}$) considered in this paper. 
Left: ranges of local density environments (Low, Intermediate  and High). 
Right: ranges of dark matter halo sizes (Small, Medium and Large).  
}\label{tablacontrol}
\centering
\begin{tabular}{cccccc}
\hline 
$Environment$  & $\Sigma \,\,\,(\rm Mpc^{-2} h^{-2}) $    & & & $Halo Mass$ & $\rm M_{vir}  \,\,\, (\rm 10^{10}\, M_{\odot} h^{-1})$ \\
\hline\hline
$Low$          & $\log \Sigma < -0.57$                    & & & $Small$     & $0$  \\       
\hline
$Intermediate$ & $-0.57 <\log \Sigma <0.05$               & & & $Medium$    & $0<\rm M_{vir}<13.5$  \\                 
\hline
$High$         & $\log \Sigma > 0.05$                     & & & $Large$     & $\rm M_{vir}>13.5$  \\              
\hline
\end{tabular}
\end{table*}

\begin{figure}
\centering
\includegraphics[width=6cm,height=5cm]{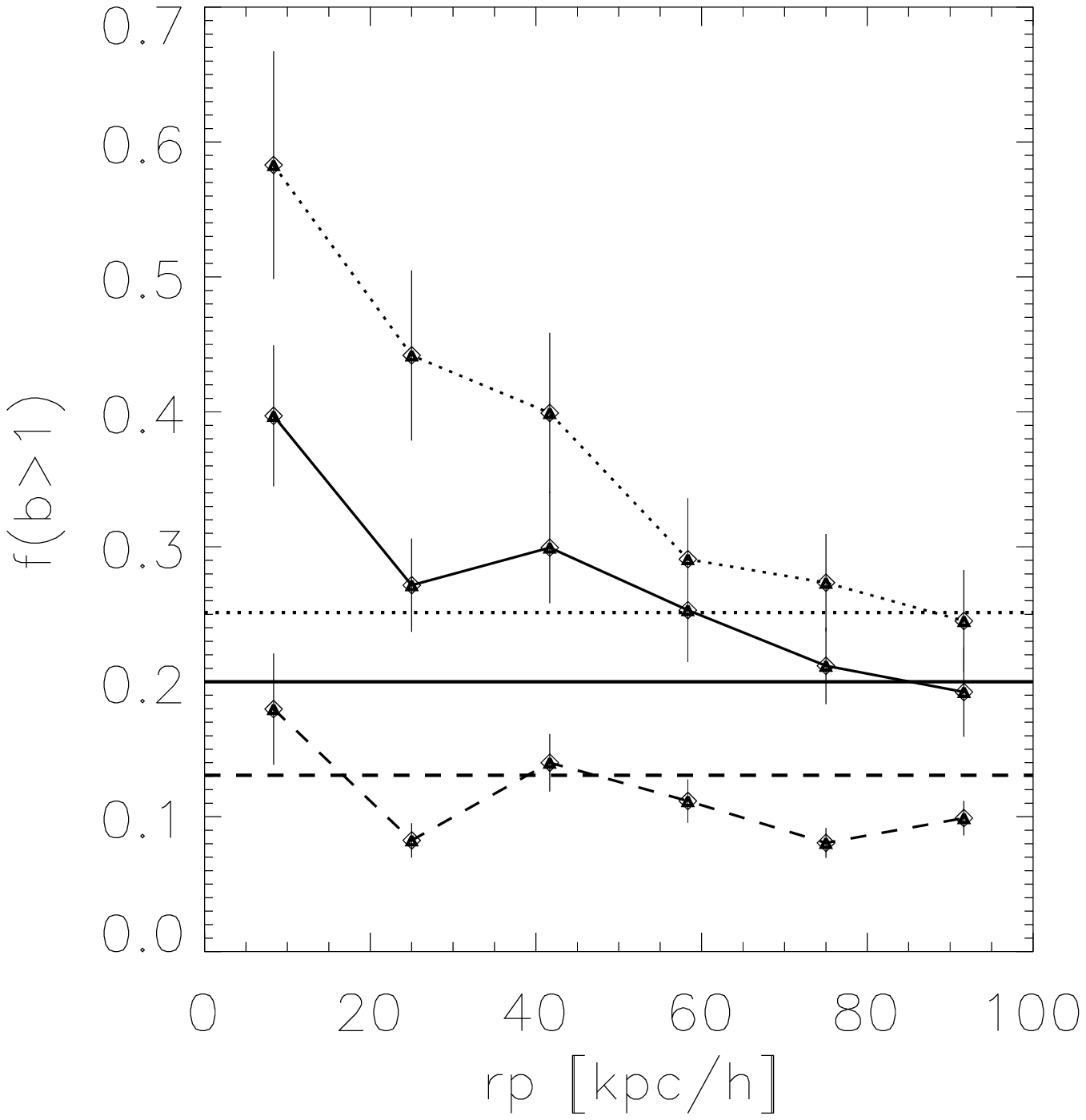}
\includegraphics[width=6cm,height=5cm]{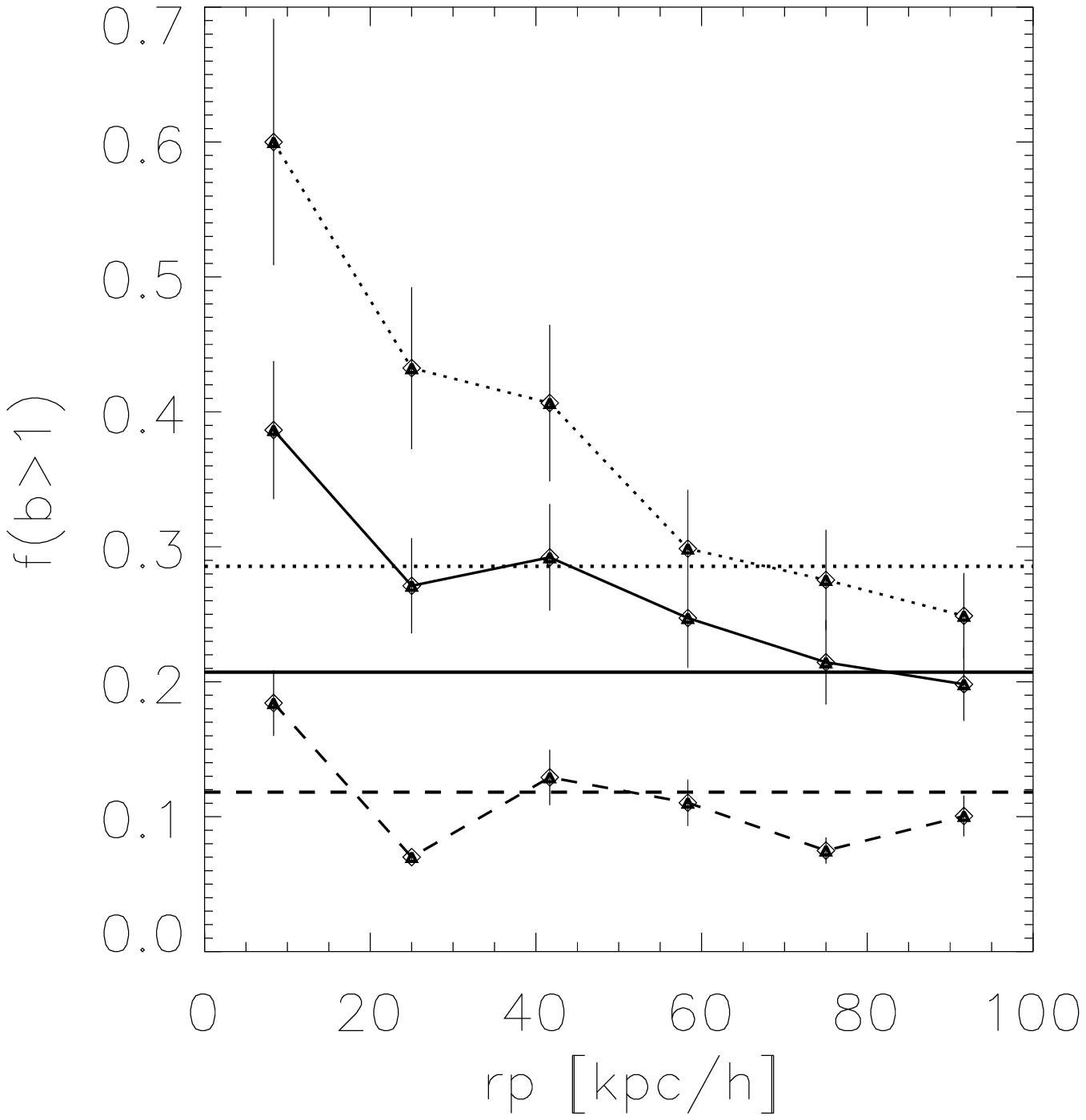}
\caption{Fraction of strong star forming galaxies ($b>1$ where $b$ has been normalized to the mean $b$ of the corresponding control sample)  in the GP catalogue
 as a function of their relative projected separation, $r_{p}$, 
at high (dashed lines), intermediate (solid lines) and low (dotted lines)
density environments (see Table 1). Horizontal lines represent the 
 fractions associated to each CS estimated with respect to their mean SF activity.
In the upper panel we show the result for  {\bf GP2} and  {\bf CS2}, and in the lower one,
for {\bf GP4} and {\bf CS4}. }
\label{b-rp}
\end{figure}

Alonso et al. (2006) have previously  analysed the efficiency of  galaxy interactions  in
driving the SF, and how the environment can modify this efficiency. 
They show that, at low and intermediate density regions, 
close galaxy interactions are more effective at triggering important SF activity 
than galaxies without a near companion. 
We revised their estimates using samples 2 ($\bf CS2$ and $\bf GP2$) and samples 4 ($\bf CS4$ and $\bf GP4$) for the purpose of assessing possible bias effects. 
Fig.~\ref{b-rp} shows the fraction of strong SF galaxy pairs 
 as a function of their relative projected separations, $r_{p}$. This fraction is defined by systems with SF activity
larger than the mean $b$ of their corresponding control sample so that they have $b >1$. The analysis was performed 
at high (dashed lines), intermediate (solid lines) and low (dotted lines)
density environments (see Table 1). Horizontal lines represent 
the mean value of the fraction associated to the corresponding CS in each environment. 
We reproduce the findings of Alonso et al. (2006)  where the only 
significant change is detected in  low density regions, where we find that galaxy pairs are required to be
closer than in {\bf GP2} (within $1 \sigma$) in order to exhibit an enhanced SF activity. 
This is a consequence of an increment of the SF activity 
in the {\bf CS4} with respect to that measured in the {\bf CS2} at low densities, indicated by 
the increase of the fraction of $b >1$ galaxies  from 0.25 to 0.29 (see also  Fig.~\ref{b-sigma} (right panels)).

Given that galaxy pairs 
closer than a critical relative  
projected separation
are more efficient at forming stars
than their isolated counterparts, 
we will concentrate our analysis on studying 
the properties of close galaxy pairs from this point on. The lower panel of Fig.~\ref{b-rp} shows that regardless of environment, systems 
 with $r_{p} < 20 \,{\rm kpc \,h^{-1}}$ have a statistically  significant enhancement
in their SF activity. Thus, we select close galaxy pairs by using this threshold on the relative 
projected separation. 

\begin{figure}
\centering
\includegraphics[width=6cm,height=5.cm]{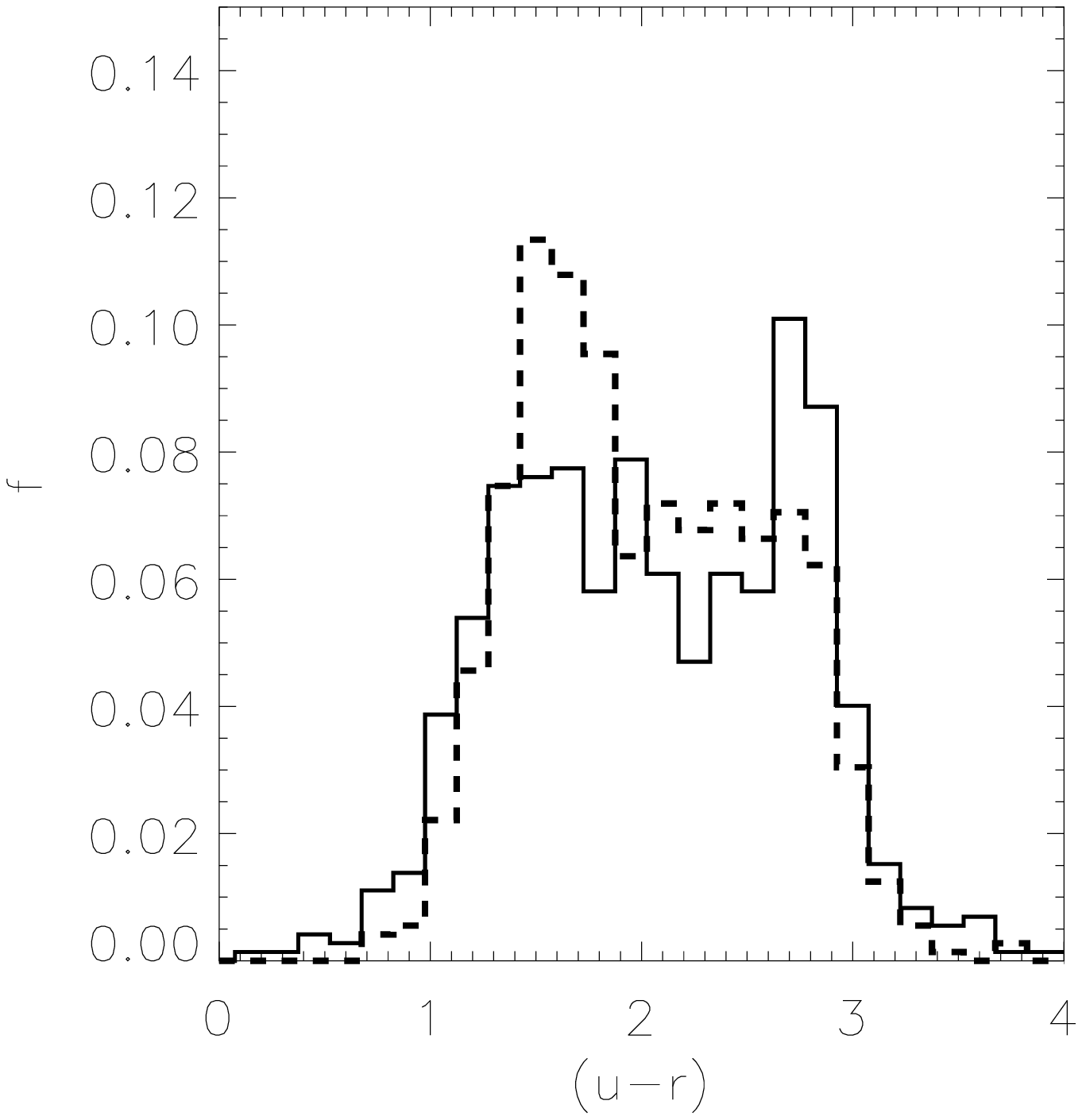}
\includegraphics[width=6cm,height=5.cm]{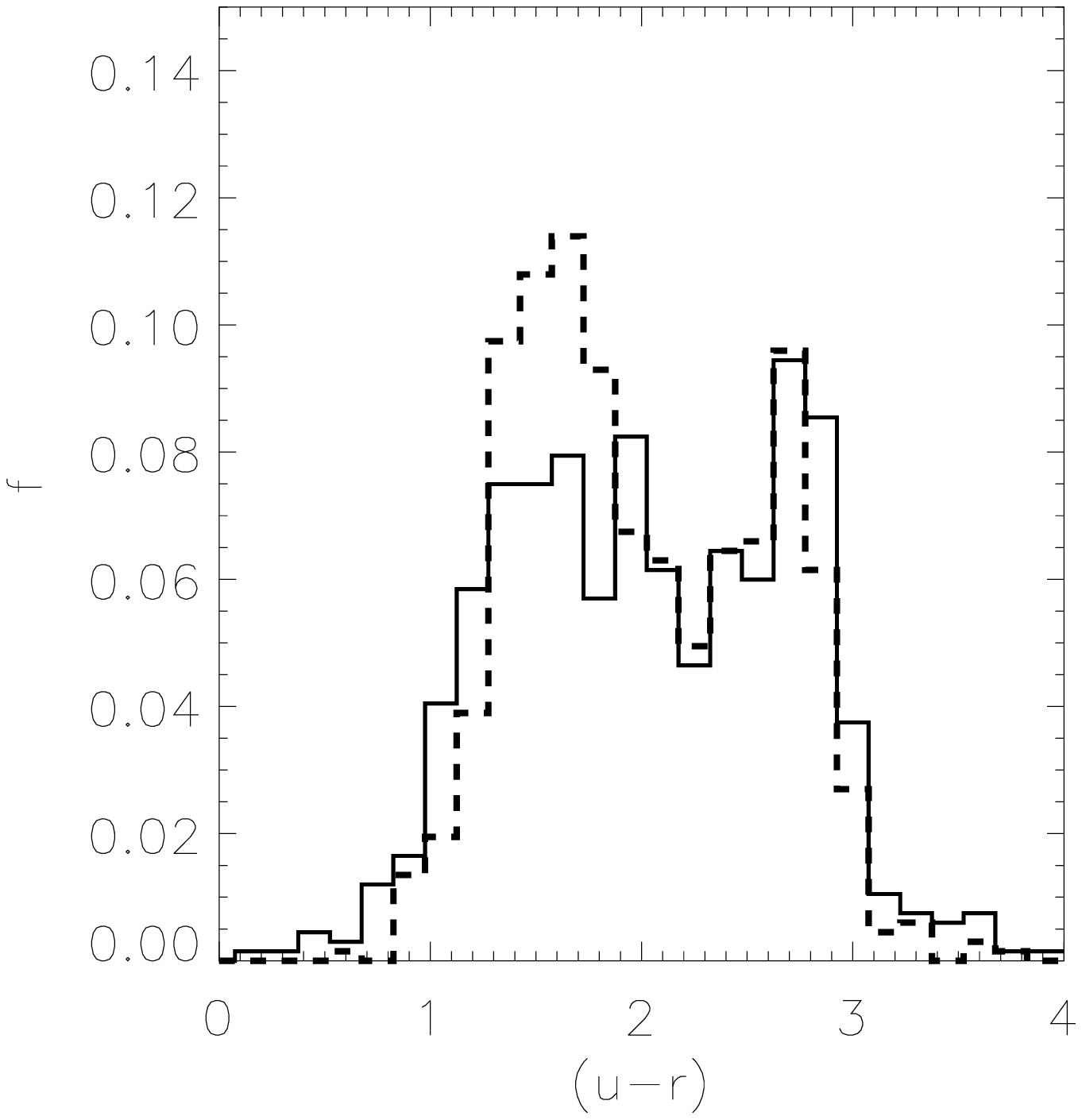}
\caption{Colour distributions  of close galaxy pairs (solid lines)
and control samples (dashed lines). 
The upper panel shows both pair and control galaxies of {\bf samples 2} and
lower panel the same for {\bf samples 4}. }
\label{color}
\end{figure}

Regarding colour distributions, observational results (e.g. Alonso et al. 2006; De Propris et al. 2005) have reported an excess of  blue galaxies in close pairs with respect to that found in their CS,
 associated with a larger fraction of actively star-forming galaxies. They also find 
a larger fraction of red galaxies in close pairs with respect to those systems without a near companion, 
an effect that might indicate that dust, stirred up during encounters, could
affect colours, partially obscuring the tidally-induced SF (Gallazzi et al. 2008). 
Other possible interpretation of this trend is that many galaxies in pairs have been very efficient at forming stars
during the early stages of their evolution, so that at present they exhibit red colours. 
In Fig.~\ref{color}, we can see that 
even though the differences at the red peak are reduced
after removing the bias effects in the selection of
the {\bf CS4}, 
the excess of close galaxy pairs in both red and blue tails 
with respect to the  {\bf CS4} still persists, supporting the claim 
that these trends are actually produced by 
galaxy interactions and  not introduced by a biased selection.

\section{Global environmental effects versus galaxy interactions.}

Studies of galaxy properties and their dependence on  environment are
key to understand the role of mechanisms driving the evolution of galaxies. 
In this section, we analyse  close galaxy interactions and mergers 
as possible environmental processes leading to evolutionary transformations. In order to
asses  how efficient galaxy interactions are with respect to other environmental mechanisms 
we analyse the properties of close galaxy pairs in different environments characterized
by their local galaxy density and by their host dark matter halo masses.

\subsection{Galaxy interactions}
\begin{figure}
\centering
\includegraphics[width=8.5cm,height=11.5cm]{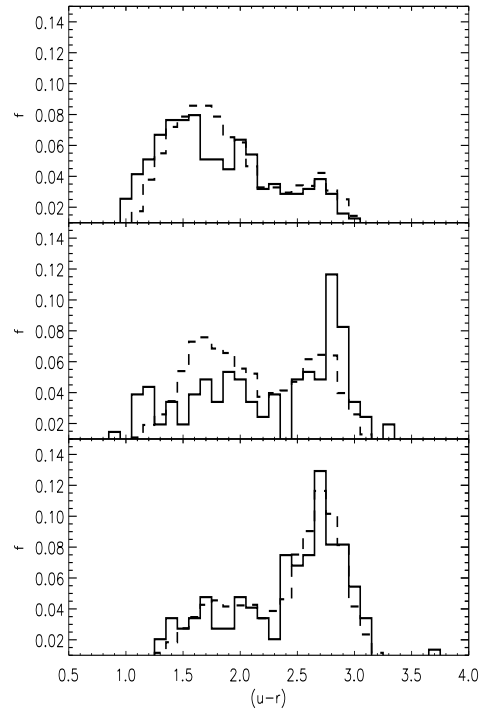}
\caption{Colour distributions of close galaxy pairs (solid lines)
and of {\bf CS4} (dashed lines), at low (upper panel), 
intermediate (middle panel) and high (lower panel) density environments.}
\label{color-sigma}
\end{figure}

The analysis of the colour distributions of galaxies by Baldry et al. (2004, also Balogh et al. 2004)  
 showed that these distributions could be well fitted by a double Gaussian over a wide
range of absolute magnitudes and local densities. The negligible variation of the blue and red peak locations of this bimodal distribution
with local galaxy density at fixed luminosities suggests that the
 process responsible for the transformation from blue to red colours needs to be very fast and efficient in order 
to overcome the effects of environment. 
In Fig.~\ref{color-sigma}, we show the $u-r$ colour distributions of close galaxy pairs 
in comparison to their isolated counterparts in {\bf CS4}. The analysis was performed for
 the three different  local density environments previously considered (see Table 1). 
We find that at  low and  high density environments (upper and lower panel, respectively), the colour distributions of
close pairs  are more similar to those of {\bf CS4} than at intermediate local densities. 
As expected, low density environments 
are  mostly populated by blue galaxies whereas high densities are populated by red galaxies. 
Interestingly, at intermediate environments (middle panel), 
the close pair and the {\bf CS4} colour distributions exhibit  significant differences, with
pairs having a larger fraction of red members.
This might indicate  that close pairs could have experienced a more efficient transition from blue to red colours,
while isolated galaxies in   {\bf CS4} experience a  more inefficient transformation. 

It is
remarkable that the more significant disagreement in their colours is observed
at both the blue and red tails, with a clear excess of 
extremely blue and red systems in close pairs (see also Alonso et al. 2006). 
The blue excess of close galaxy pairs is expected to  be tidally induced, associated to the 
 enhanced SF already shown in Fig.3 (e..g Lambas et al. 2003; Nikolic et al. 2004).
The extremely red close pairs could be accounted  for by dusty, obscured star-forming systems, 
as suggested by observations which indicate  
an intrinsic reddening associated to the near infrared emission  of 
hot dust present in  tidally-triggered star forming regions (Geller et al. 2006; Lin et al. 2007).
Interestingly, there are  observational indications of an excess of  
the red, obscured star-forming galaxies  
at intermediate densities 
(Wolf et al. 2005; Gallazzi et al. 2008; Poggianti et al. 2008).
These evidences may support the idea of a connection between 
 the excess of red close galaxy pairs at intermediate densities and a tidally induced dusty SF. 
However, there are other plausible interpretations.
Red galaxies in pairs could be systems dominated by  old stellar populations which had their gas reservoir stripped as they entered higher density regions so that interactions are now 
unable to trigger new SF events. This possibility  will be tested in the following sections when
we analyse the dependence on environment since this should  affect both galaxies in pairs and in
isolation.  Alternatively, isolated galaxies could also include merger remnants
with blue colours as a consequence of the recently triggered SF activity while  close
galaxy pairs could be just starting to interact and, hence, reflect only the colour of the underlying 
stellar population.  Unfortunately, we cannot address this possibility using the currently available data.

\begin{figure}
\centering
\includegraphics[width=8.5cm,height=11.5cm]{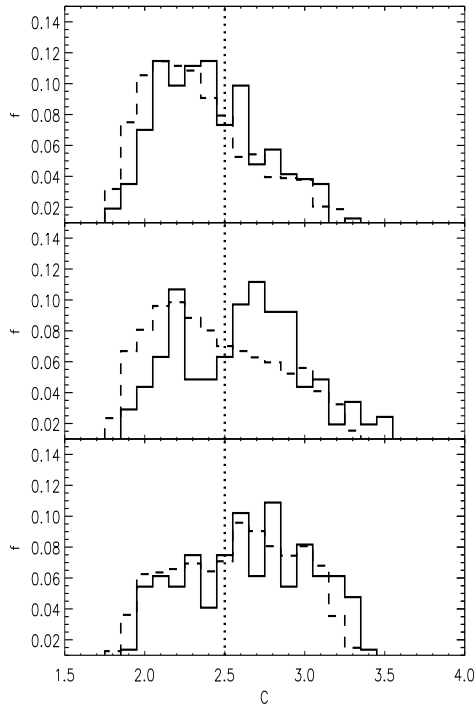}
\caption{Concentration index ($C=r90/r50$) distributions
of close galaxy pairs (solid lines) and of {\bf CS4} (dashed lines), for low (upper panel), 
intermediate (middle panel) and high (lower panel) density environments.
Vertical dotted lines represent the critical concentration index value  ($C=2.5$)
 adopted to segregate bulge ($C >2.5$) from disc ($C<2.5$) dominated morphologies.}
\label{C}
\end{figure}

Even when the morphological classification of interacting systems 
is  difficult to perform, especially in the case of mergers and 
close galaxy pairs with tidal distortions,  one can consider the concentration index, $C$, defined 
by the ratio of the Petrosian $90\% - 50\%$ r-band light radii, as
a global indicator of the structural luminosity distribution in these galaxies. 
Bearing this in mind, we explore the concentration index distributions
of close galaxy pairs in comparison with those of {\bf CS4} 
at the same three different  local density environments of Fig.~\ref{color-sigma}.  
In order to highlight the result of  Fig.~\ref{C}, we plot a dotted vertical line indicating
 the critical concentration index value of $C=2.5$  
 adopted to segregate concentrated,  bulge-like ($C >2.5$) galaxies from more extended,  disc-like ($C<2.5$)  systems.  
The concentration index distributions of close pairs approximately match those of {\bf CS4}
at low and high density environments,
following the expected morphology-density relation (Dressler 1980; Gomez et al. 2003), with 
a larger fraction of disc (bulge) dominated systems populating the low (high) density environments.  
Again it is at  intermediate density regions where we find a significant difference, indicating that
 most of the members of pairs tend to be bulge-dominated systems.

The similar trends found  for colours and morphologies as a function of environment 
(Fig.~\ref{color-sigma} and Fig.~\ref{C})  suggest   a common physical mechanism 
 responsible for both the galaxy colour and  the morphological transformations. 
     
\subsection{Global environmental effects}
Up to this point,  we have only analysed the role of close galaxy interactions
at different local density environments. However, many other environmental processes
(e.g., ram-pressure stripping, starvation, harassment, etc.) may collaborate to produce
the trend found at intermediate densities. In order to isolate the  effect of galaxy interactions 
from other environmental processes, we have also investigated galaxy properties taking
into account the masses of their dark matter host haloes. 
For this purpose, we resorted to the dark matter halo catalogues of Zapata et al. (2009) which 
identified groups in the SDSS-DR4 (with a minimum number of 10 members) by using dynamical criteria as explained in
details by the authors. Then, we
assigned each galaxy in our catalogue to one of these groups by using the criteria  explained in Section 2.
Galaxies which did not satisfy them are considered to inhabit haloes with masses lower than
$10^{13} h^{-1} M_{\odot}$.

\begin{figure}
\centering
\includegraphics[width=6.cm,height=5cm]{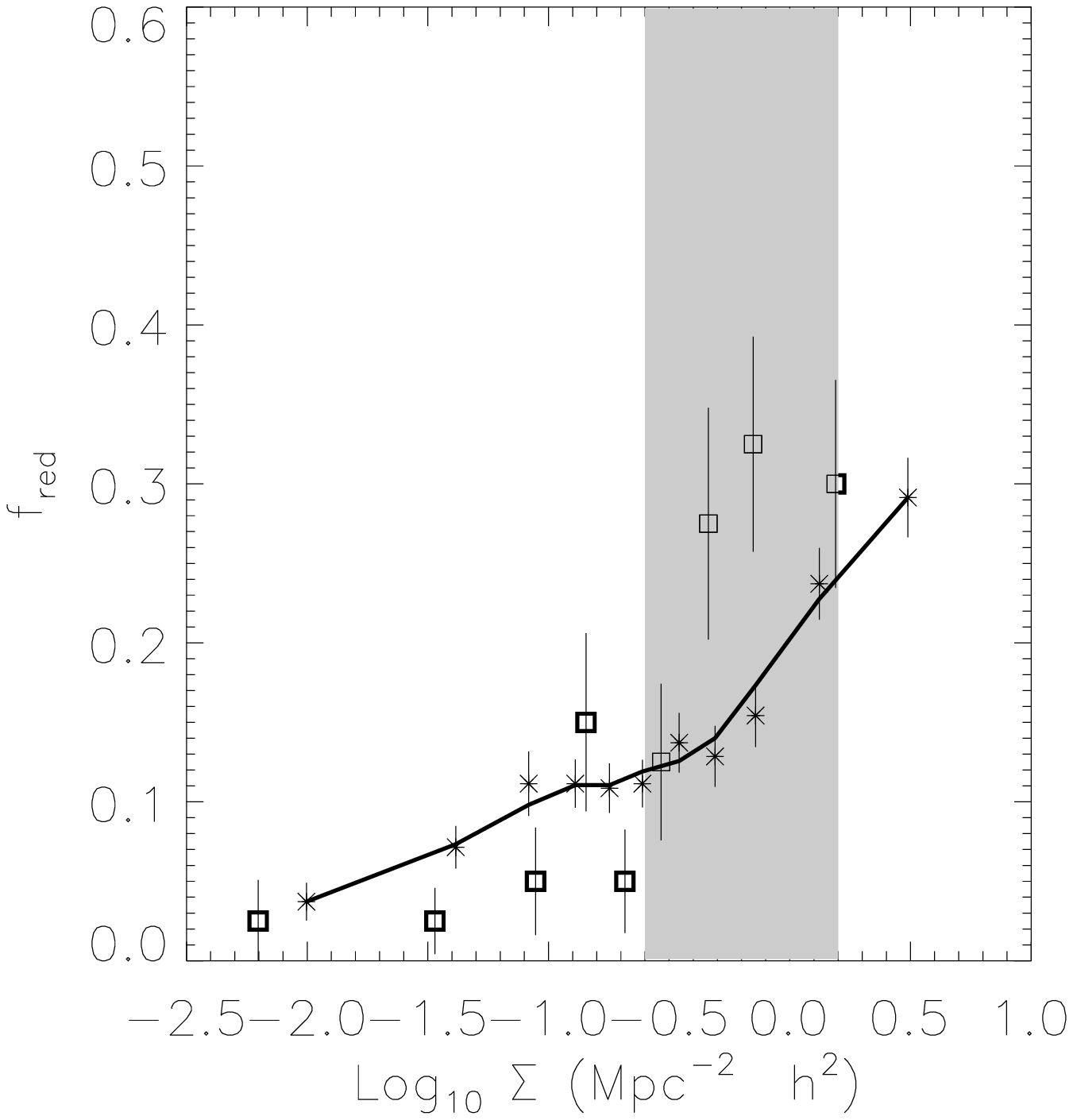}
\includegraphics[width=6.cm,height=5cm]{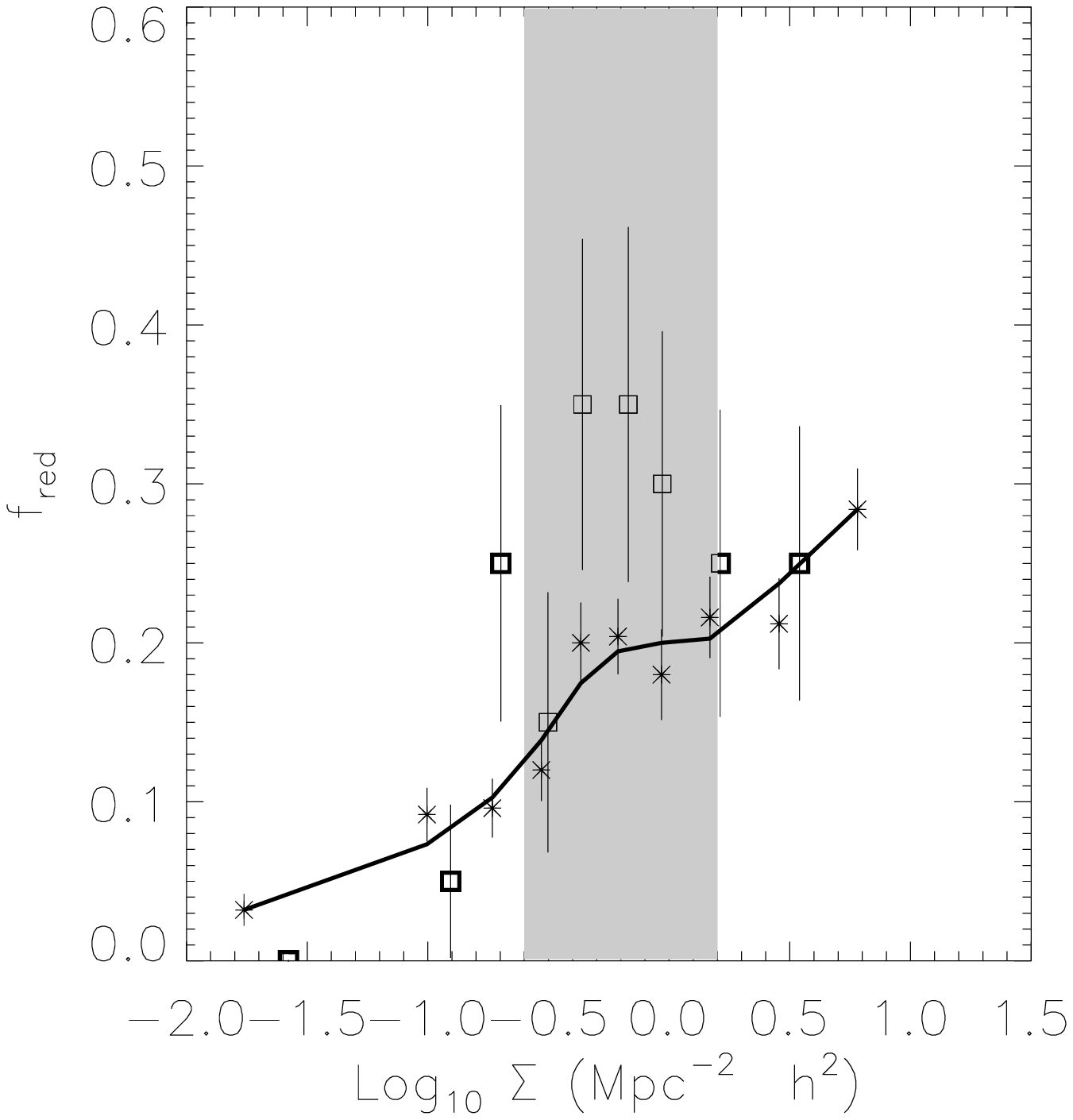}
\includegraphics[width=6.cm,height=5cm]{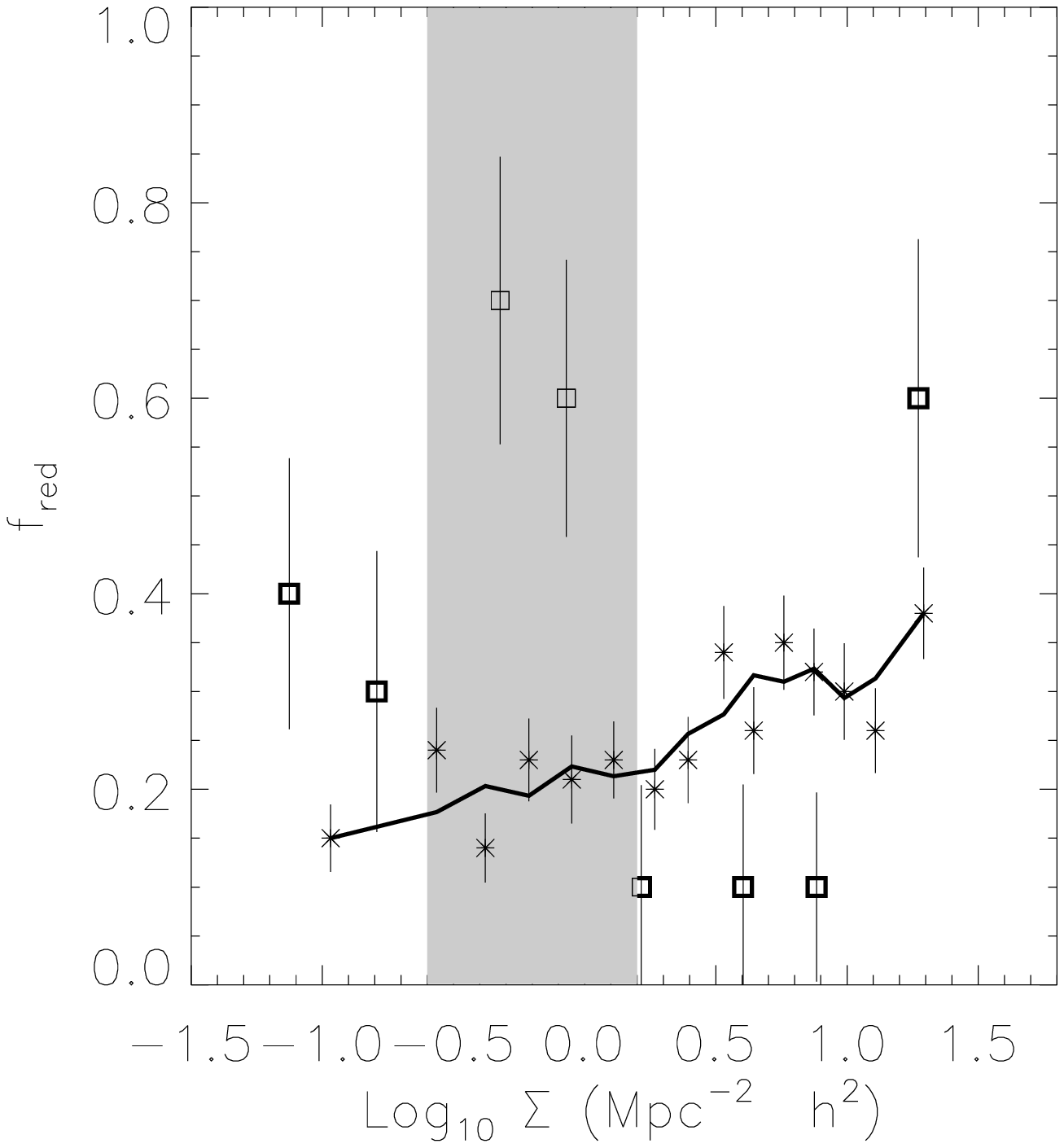}
\caption{Fraction of red systems ($(u-r)>2.5$) in close galaxy pairs (open squares)
and in the {\bf CS4} (asterisks) as a function of the local density parameter, $\Sigma$. 
Error bars represent boostrap errors and solid lines  smoothed functions of the {\bf CS4} trends. 
Shaded areas remark the intermediate local density environments.
Samples were divided in three different bins of halo masses,  from top to bottom panels: small, medium and 
large haloes (See Table 1 for descriptions).}
\label{fred}
\end{figure}

Fig.~\ref{fred} shows the red fraction ($(u-r)>2.5$) 
of close galaxy pairs and of those galaxies in the  {\bf CS4} as a function of the local galaxy density parameter  ($\Sigma$). The 
analysis was performed for  three different halo mass ranges  described in Table 1: 
small, medium and large mass haloes. 
We studied galaxies in both {\bf GP4} and {\bf CS4} which reside in the same local density regions and in haloes of similar mass and hence,
any possible signal should be presumably ascribed to galaxy interactions.
From this figure we can see that the underlying trend is consistent with the expected increase of 
the red fraction with local galaxy density for both galaxies in pairs and without a companion, which
might be associated with the action of host-halo mass dependent processes.
Interestingly, at  intermediate local densities (shaded regions), regardless
of  halo mass, the red fraction of galaxies  in close pairs (square symbols) 
comfortably exceeds that of isolated galaxies in the {\bf CS4} (solid lines), suggesting that the mechanism which drives the 
evolution of galaxies in close pairs 
at intermediate local density regions
is not regulated by host-halo mass dependent processes alone.

To quantify these results, we also calculated the percentages of red galaxies in close pairs and in the {\bf CS4} 
for the three different bins of local density environments and halo masses defined in Table 1. 
We find that, independently of their halo masses, 
at low and high local density environments, the percentages of red systems in  close pairs 
are very similar to those found in the {\bf CS4}  (columns 1 and 3 of Table~\ref{tablacontrol}). 
Computing 
the difference of percentage  between the close pairs and the {\bf CS4}
for these two columns,  we get 
a maximum value  of  $\sim 8\%$, with an increase of $\approx 10\%$ from low to high mass haloes 
for both galaxies in pairs and in the control sample. However, at intermediate densities (column 2), the gap between  the
red fraction of  close pairs and  {\bf CS4} increases remarkably from 
 $7.7\%$ in low mass haloes to  $45.6\%$ in high mass haloes.

\begin{table*}
\caption{Percentages of red galaxies, $(u-r)>2.5$, at {\em Low}, {\em Intermediate}  and {\em High} local density environments,  
hosted by  {\em Small}, {\em Medium}  and {\em Large} halo masses (see Table 1 for descriptions).
These percentage are provided for close galaxy pairs (CP) and for the {\bf CS4}.
}\label{tablacontrol}
\centering
\begin{tabular}{ccccc}
\hline 
$M_{vir}\setminus\Sigma$             & $Low$     & $Intermediate$    & $High$    & $Samples \,4$\\
\hline\hline
$Small$        & $16.9\%$  &  $37.5\%$   &  $66.1\%$ & CP \\
               & $18.9\%$  &  $29.8\%$   &  $57.9\%$ & CS4  \\
\hline
$Medium$       & $17.8\%$  &  $50.7\%$   &  $42.0\%$ & CP \\
               & $20.8\%$  &  $36.7\%$   &  $48.2\%$ & CS4  \\
\hline
$Large$        & $33.3\%$  &  $85.7\%$   &  $56.1\%$ & CP \\
               & $32.8\%$  &  $40.1\%$   &  $60.9\%$ & CS4  \\
\hline
\end{tabular}
\end{table*}

\begin{figure}
\centering
\includegraphics[width=6cm,height=5cm]{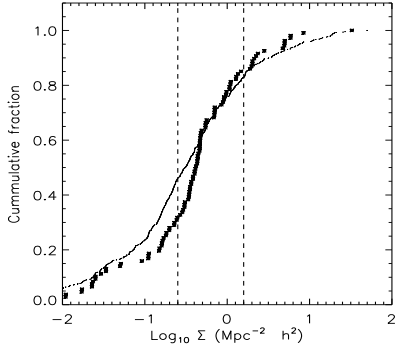}
\caption{Cummulative fraction of galaxies in close pairs (small dots) 
 and merging systems (asterisks) as a function of local density parameter. Vertical dashed
lines indicate the intermediate local density regions (see Table 1).
 }
\label{number}
\end{figure}

\begin{figure}
\includegraphics[width=8.5cm,height=12cm]{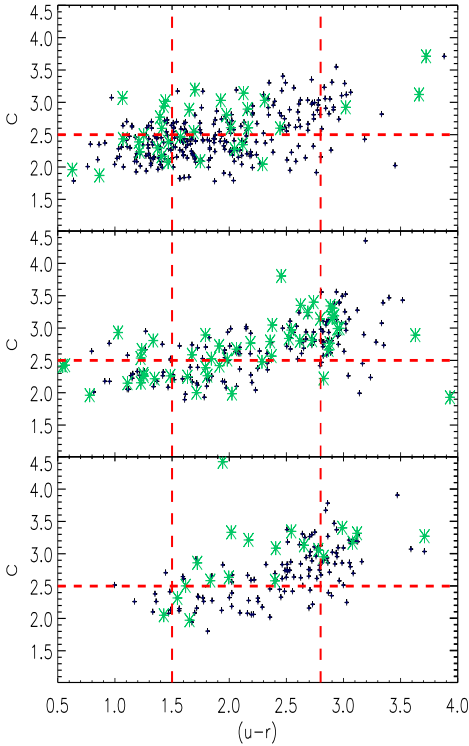}
\caption{Scatter plot of colour and concentration index for close galaxy pairs (blue small dots)
 and merging systems (green asterisks)  at low (upper panel), intermediate (middle panel) 
and high (lower panel) density environments (see Table 1). Red horizontal lines (at $C=2.5$) 
separate disc from bulge dominated systems, and vertical ones 
indicate the blue  ($(u-r)<1.5$) and red ($(u-r)>2.8$) tails in the colour distributions.
 }
\label{merging}
\end{figure}

 We have also analysed the role of mergers in different environments.
 The merging galaxy sample considered in this paper was extracted from the
 close galaxy pair sample by selecting objects with
 morphological disturbances
 and strong signals of interactions, as explained by Alonso et al. (2007).
 First, we analyse the cummulative distribution of merging systems and close galaxy pairs
 in different local density environments
 (Figure ~\ref{number}).
 As can be seen, the number of close galaxy
 pairs (small dots) and
 merging systems (asterisks) increases rapidly in intermediate density environments
 (delimited by the vertical dashed lines in the figure). This result agrees with
 previous works indicating that merging systems and galaxy-galaxy
 encounters are
 frequent in intermediate density regions (e.g. Moss 2006).
 As the figure shows,
 we find $\sim 30\%$ of close galaxy pairs and $50\%$ for merging systems at intermediate
local density environments.
 In Fig.~\ref{merging}, we show colours and morphologies of close galaxy
 pairs (blue dots) and
 merging systems (green asterisks), at the same local density bins
 previously considered.
 A large fraction of merging systems populating the sequence of
 extremely red and bulge-dominated galaxies
 can be found at intermediate local densities (middle panel).
 Particularly, $26\%$ of merging galaxies at intermediate densities are in
 the red ($(u-r)>2.8$) tail,
 a percentage which rises to $50\%$ by adding the blue ($(u-r)<1.5$)
 tail.


These
results could be interpreted within the  theoretical work of Kapferer et al. (2008) if we assume that dust obscuration is 
hiding  much of the enhanced SF activity triggered by interactions.
Kapferer et al. (2008)  find that the global SFR of 
interacting systems is largely enhanced in the presence of a moderate ram pressure, 
in comparison to the same interaction without the presence of an ambient medium. They
use hydrodynamical simulations to model interacting galaxies moving through a hot, thin medium. 
This model mimics an average low ram pressure in the outskirts of galaxy cluster 
where systems interact at low velocities, resembling  galaxy interactions occurring within groups,
falling  along cluster filaments.  Combining this prediction with the observational results that show an increment 
of red dusty star forming systems at intermediate environments (Gallazzi et al. 2008), 
we might suggest that, at intermediate density regions, where galaxy interactions
are more frequent, low ram pressure stripping from the diffuse intra-group medium could  collaborate to 
enhance the effect of galaxy interactions,  rising the
fraction of red star-forming  galaxies  in close pairs up to almost  $50\%$ over that found in the {\bf CS4}. 

Nevertheless, we should warn about the fact that the large excess 
of red galaxy fractions in  close pairs 
at intermediate local densities and within large halos, 
could be overestimated. Large haloes tend to have larger differences between the central and satellite galaxy luminosities
than smaller ones. Consequently, some galaxies could be mistakenly classified
as control samples although they could have a near companion, faint enough to be undetected by
the SDDS.

\section{Summary}

Recently, many authors have proposed that intermediate density regions (i.e. galaxy groups in the outskirt of clusters,  
infalling populations) are the locations where the local environment influences the transition of galaxies onto the
red-sequence, as opposed to mechanisms that operate on cluster scales. 
In this paper, we use the SDSS-DR4 data to explore the role of mergers 
and interactions as  environmental processes driving evolutionary 
transformations of galaxies. 
We analyse properties of galaxy pairs and merging systems 
in different local density environments and with different host halo masses, 
comparing these properties with those of isolated galaxies in an unbiased CS.

We build a Galaxy Pair Catalogue from SDSS-DR4  data, 
by requiring its members to have relative projected separations: $r_{p}< 100 \rm \,kpc \,h^{-1}$
and relative radial velocities: $\Delta V< 350 \rm \,km \,s^{-1}$. 
For comparison, and 
in order to isolate the effects of galaxy interactions, we have considered
previous theoretical findings (P09) 
and build an 'unbiased' CS imposing constrains 
on redshift, stellar mass, local density and halo
mass, to match those of galaxies in the Pair Catalogue.
The analysis of biases in the selections of the SDSS CS shows that, contrary to semi-analytical simulations, 
the local density environment is responsible for introducing the largest bias effect, 
giving  the dark matter halo mass a less significant role in the CS definition 
than previously reported (Barton et al. 2007; P09). Even when  the dynamical method  to estimate
halo masses is still crude, 
this finding might be an indication of an unsuitable treatment of satellite galaxies in the SAMs. 

In order to asses the efficiency of galaxy interactions  in comparison
to other environmental mechanisms,
we analyse the properties of close galaxy pairs in different local 
environments and residing in host haloes of different masses, 
in comparison to isolated galaxies in the 'unbiased' control sample in similar environments.
We have characterized the local environment by means of the local galaxy projected density, 
computed with the $5^{th}$ nearest bright neighbour, 
and via constraints on their host DM halo masses. 

Our results can be summarized as follows:

(i) We analyse the colour distributions of close galaxy pairs in comparison to those of the 
control sample 
at low, intermediate and high local densities. We find that, at  low and high density environments, 
the colour distributions of close pairs are similar to those of the control sample. 
An opposite effect is found at intermediate densities, 
where  their colour distributions exhibit  significant differences, 
indicating that close pairs could have experienced a more rapid transition from blue to red colours,
while the isolated galaxies in the control sample undertake a  more inefficient transformation.  
The significant disagreement in their colour distributions is detected 
in  both colour tails, with a clear excess of extremely blue and red systems in close pairs.  
We speculate that  these excesses might be likely  associated with the enhancement of obscured (e.g. Gallazzi et al. 2008) and unobscured (e.g. Alonso et al. 2006) SF activity tidally induced by close interactions.  
Other explanations are possible as well, as is discussed in Section 3.1

(ii) In consistence with our analysis on colours, the distributions of concentration indexes of
close galaxy pairs and galaxies in the control samples shows that they are similar at  low and high local density environments, 
following the expected morphology-density relation. Again, it is at intermediate local density regions 
where we find important differences, indicating that  most of the galaxies in pairs are  bulge-dominated systems.  The correlation found  between colours and morphologies suggests that 
the physical mechanism responsible for the colour properties could also 
operate transforming the galaxy morphologies.

(iii) We find that $\sim 50\%$ of the merging systems visually identified in our close galaxy pair catalogue  inhabit  intermediate local densities. We detect that, at this intermediate densities, a large
 fraction of mergers are  red 
and bulge dominated. Particularly, $26\%$ of these mergers are in the red ($(u-r)>2.8$) tail while $24\%$ are located at the blue ($(u-r)<1.5$) extreme.

(iv) We have also investigated galaxy interactions at different halo masses 
in order to isolate the effects coming from this environmental indicator.
For low and high local densities, we detect an increase in the red fraction for increasing
dark matter halo mass as expected, and with similar levels for galaxies in pairs and in the control sample.
However, at  intermediate local densities and regardless of halo mass, 
we find that the  red fraction of systems in close pairs largely exceeds 
that of isolated systems in the control sample, up to $\sim 50\%$ for the largest haloes.
Under the hypothesis that the red colours are produced by obscured SF due to  dust stirring,
triggered by the interaction, our  results would be consistent with  those of Gallazzi et al. (2008) who actually found an excess of obscured star formation at intermediate local densities.  
From a theoretical point  of view, the larger excess at intermediate local densities could be interpreted by the findings of Kapferer et al. (2008) who 
claimed that the global SF rate of interacting systems  is largely enhanced 
by a moderate ram pressure stripping, in comparison to the same interaction without the presence of an ambient medium. 

We conclude that mergers and galaxy interactions are important processes in the regulation of 
galaxy properties, particularly at intermediate local density environments where these interactions 
seem to be more  frequent. Although we detect the action of the host DM halo  in the steady increase of the
red fraction as a function of halo mass, we also unveil a distinctive behaviour for galaxies in pairs at intermediate density environments. These regions could be considered as the locations where galaxies are pre-processed by mergers and close encounters that transform their colours and morphologies.
Finally, we suggest that our  findings could also help to improve the modelling used in SAMS,
particularly regarding the effects of galaxy-galaxy interactions which are 
currently not included in most semi-analytic schemes.

\section*{Acknowledgments}
This work was partially supported by the
Consejo Nacional de Investigaciones Cient\'{\i}ficas y T\'ecnicas  (PIP 6446) 
and Agencia Nacional de Promoci\'on Cient\'\i fica y T\'ecnica (PICT 32342 (2005) and PICT Max Planck 245 (2006).


\label{lastpage}
\end{document}